\documentclass[showpacs,epsf,onecolumn]{revtex4-1}
\usepackage{float}
\usepackage{color,graphicx}
\usepackage{hyperref}
\begin{document}

\title{Coexistence of topological Dirac fermions in the surface and three-dimensional Dirac cone state in the bulk of ZrTe$_{5}$ single crystal}
\author{Arnab Pariari \& Prabhat Mandal}
\address{Saha Institute of Nuclear Physics, 1/AF Bidhannagar, Calcutta 700 064, India}
\date{\today}
\begin{abstract}
 {\large Although, the long-standing debate on the resistivity anomaly in ZrTe$_{5}$ somewhat comes to an end, the exact topological nature of the electronic band structure remains elusive till today. Theoretical calculations  predicted that bulk ZrTe$_{5}$  to be either a weak or a strong three-dimensional (3D) topological insulator. However, the angle resolved photoemission spectroscopy and transport measurements clearly demonstrate 3D Dirac cone state with a small mass gap between the valence band and conduction band in the bulk. From  the magnetization and magneto-transport measurements on ZrTe$_{5}$ single crystal, we have detected both the signature of helical spin texture from topological surface state and chiral anomaly associated with the 3D Dirac cone state in the bulk. This implies that ZrTe$_{5}$ is a novel 3D topological insulator having massless Dirac fermionic excitation in its bulk gap state. Whereas, no 3D topological insulator known in material science holds linear band dispersion in its insulating bulk. Apart from the band topology, it is also apparent from the resistivity and Hall measurements that the anomalous peak in the resistivity can be shifted to a much lower temperature ($T$$<$2 K) by controlling impurity and defects.}\\
\end{abstract}
\maketitle
The low-dimensional pentatellurides, ZrTe$_{5}$ and HfTe$_{5}$, synthesized in 1973 [1], exhibit a  peak in the resistivity ($\rho$) as a function of temperature [2]. This anomaly in the resistivity has been observed at T$_{P}$ $\sim$ 145 K for ZrTe$_{5}$ and T$_{P}$ $\sim$ 80 K for HfTe$_{5}$, however,  the exact temperature varies from sample to sample depending on the impurity level [3]. With decreasing impurity, the anomalous peak in resistivity shifts to lower temperature [3]. Recent works on ZrTe$_{5}$ reported T$_{P}$ as low as $\sim$60 K, which has been ascribed to very low defect and impurity concentration in the samples [4, 5]. Most of the earlier works have been directed towards understanding the origin of this peak. The charge carrier switches from holes at $T$$>$T$_{P}$ to electrons for $T$$<$T$_{P}$, which is reflected in the sign change of thermoelectric power [6] and Hall coefficient [7]. Initially, it was believed that this resistive anomaly arises due to a charge-density wave  transition, similar to that  occurs in NbSe$_{3}$ [8]. But the absence of lattice modulation, etc., eliminate the idea of charge density wave formation in ZrTe$_{5}$ [9].  Subsequently, the concept of possible polaronic conduction [10], semimetal-semiconductor phase transition [11] and so on, have emerged until a recent theoretical work suggests that the single layer of ZrTe$_{5}$ and HfTe$_{5}$ crystal is the most promising candidate for the quantum spin Hall due to the large bulk gap [12]. Suddenly, a material known for its large thermoelectric power [6], resistivity anomaly [1] and large positive magnetoresistance [13], has been brought to our attention to study the topological properties of relativistic Dirac fermion in condensed matter physics [4, 5, 14-21]. It has been established from recent angle resolved photoelectron spectroscopy (ARPES) measurement that the temperature dependence of the electronic band structure across the Fermi energy is responsible for the anomalous peak in resistivity [14]. However, ZrTe$_{5}$ is not free from debate, facing more bigger question. Theoretical calculation shows that electronic properties of bulk ZrTe$_{5}$  is very sensitive to the lattice parameters. Depending on the values of lattice parameters it can be either a weak or a strong three-dimensional topological insulator [12]. On the other hand,   ARPES [4, 14], infrared spectroscopy [5, 15] and magneto transport [4] studies show three-dimensional linear dispersion with a small semiconducting gap  between valence and conduction band, i.e., 3D Dirac fermionic excitation with small mass gap. Do theory and experiment contradict with each other or the topological Dirac fermions in the surface and three-dimensional Dirac cone state in the bulk can coexist simultaneously  in ZrTe$_{5}$? If the later is  possible, it would be remarkable phenomenon. We will have a three-dimensional topological insulator with Dirac fermionic excitation in its bulk.\\
{\textbf{\large Results}}\\
{\textbf{Crystal structure.}}
High quality single crystals of ZrTe$_{5}$ were grown by iodine vapor transport method similar to that  reported earlier [22]. Typical size and morphology of few representative single crystals are shown in Fig. 1(a). Phase purity and the structural analysis of the samples were done by high resolution powder x-ray diffraction (XRD) technique (Rigaku, TTRAX II) using Cu-K$_{\alpha}$ radiation [see Supplementary Figure 1]. Within the resolution of XRD, we have not seen any peak due to the impurity phase. The calculated value of the lattice parameters are $a$=3.96 {\AA}, $b$=14.50 {\AA} and $c$$=$13.78 {\AA} with space group symmetry $Cmcm$, similar to the earlier reports [23-25]. The structure of the pentatellurides consists of trigonal prismatic chains of ZrTe$_{3}$ along \textbf{a} axis that linked via parallel zigzag chains of Te atoms along the \textbf{c} axis, which together form 2D planes weakly bonded via van der Waals force along the \textbf{b} axis [12]. Figure 1(b) shows the crystallographic directions of a typical ZrTe$_{5}$ single crystal.\\
{\textbf{Temperature dependence of resistivity both in presence and absence of external magnetic field.}}
Resistivity and transverse magnetoresistance measurements are done by applying current along the \textbf{a} axis and magnetic field perpendicular to the \textbf{ac} plane, i.e., along \textbf{b} axis. Figure 1(c) shows the temperature dependence of resistivity of ZrTe$_{5}$ single crystal both in presence and absence of  magnetic field. The zero-field $\rho$ exhibits metallic behavior (d$\rho$/d$T$$>$0) down to 25 K. Below 25 K, $\rho$ shows a weak upturn, i.e., a crossover from metallic to semiconducting like behavior. However, several earlier reports show that a broad peak appears in the temperature dependence of $\rho$, which is known as the resistivity anomaly of ZrTe$_{5}$ [1-5]. We have already mentioned  that the temperature at which $\rho$ shows peak varies widely; from 60 K to 170 K depending on the presence of impurity and defect concentration in the sample. It has been argued that the binding energy shift of valence and conduction bands as a function of temperature is responsible for the peak at $T_{P}$ [14]. The sign of the charge carrier changes from positive (hole) to negative (electron) and the peak in $\rho$($T$) appears when the chemical potential crosses the gap ($\sim$50$\pm$10 meV) from valence band to conduction band. The absence of resistivity peak down to 2 K in the present sample could be attributed to much smaller impurities and defects. Under application of magnetic field, $\rho$ increases sharply at low temperature and the metal-semiconductor crossover shifts to higher temperature, which is consistent with the earlier reports [4, 25]. But, no re-entrant metallic state has been observed up to 9 T.\\
\begin{figure}
\includegraphics[width=1.0\textwidth]{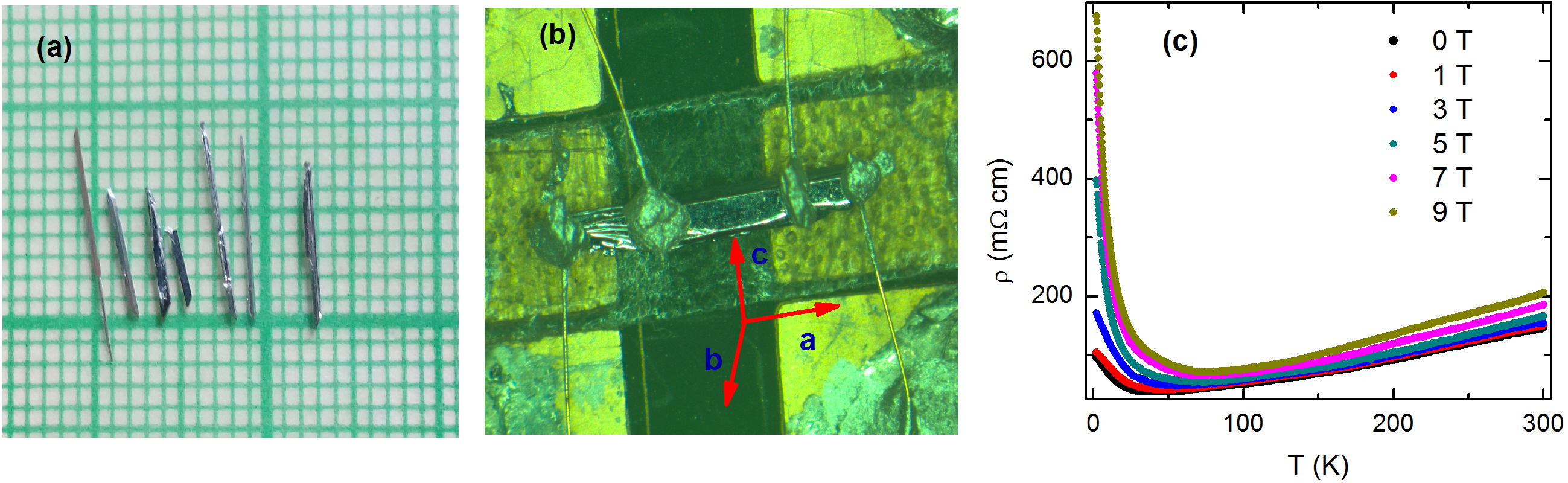}
\caption{(Color online) (a) Typical size and morphology of few representative single crystals of ZrTe$_{5}$, (b) Different crystallographic directions, and (c) Temperature dependence of resistivity ($\rho$) both in presence and absence of external magnetic field.}\label{fig1}
\end{figure}
{\textbf{Hall resistivity and transverse magnetoresistance.}}
To further ensure the absence of resistivity anomaly, which has been ascribed to the switching of $p$-type semimetal to $n$-type semimetal state, we have done Hall measurements down to 2 K.  Figure 2(a) shows that the Hall resistivity ($\rho_{xy}$) remains  positive over the entire temperature range 2-300 K. The absence of sign change in $\rho_{xy}$ is consistent with the observed $T$ dependence of $\rho$.  One can see that $\rho_{xy}$ is linear over the entire field range except at low temperature, where  an upward curvature appears at high fields due to the Shubnikov-de Haas oscillations. A systematic increase of the slope of the Hall resistivity with decreasing temperature is consistent with the temperature evolution of electronic band structure in ZrTe$_{5}$ [14, 21]. From the slope of $\rho_{xy}$($H$), the bulk carrier density ($n$) is calculated to be $\sim$4$\times$10$^{16}$cm$^{-3}$ and $\sim$7$\times$10$^{16}$cm$^{-3}$ at 2  and 300 K, respectively. We would like mention that the carrier density in the present crystal is almost one order of magnitude smaller than the earlier reported ones [18, 26]. Figure 2(b) shows the normalized magnetoresistance (MR) up to 9 T magnetic field.  MR is large, positive and shows no sign of saturation in the measured temperature and field range. The observed behavior of MR is similar to the earlier reports [[13, 16, 17]. At low temperature, the MR is dominated by a very low frequency ($\sim$3 T) Shubnikov-de Haas oscillation, which implies the presence of a tiny Fermi pocket, consistent with the low carrier density determined from the Hall measurements. Employing the Onsager relation
$F$$=$($\phi$$_0$/2$\pi$$^2$)$A_F$, we have calculated the cross-sectional area ($A_F$) of the Fermi surface normal to the field $\sim$6.2$\times$10$^{-5}$${\AA}^{-2}$. At high temperatures, where the quantum oscillation suppresses, MR becomes linear.\\

\begin{figure}
\includegraphics[width=0.8\textwidth]{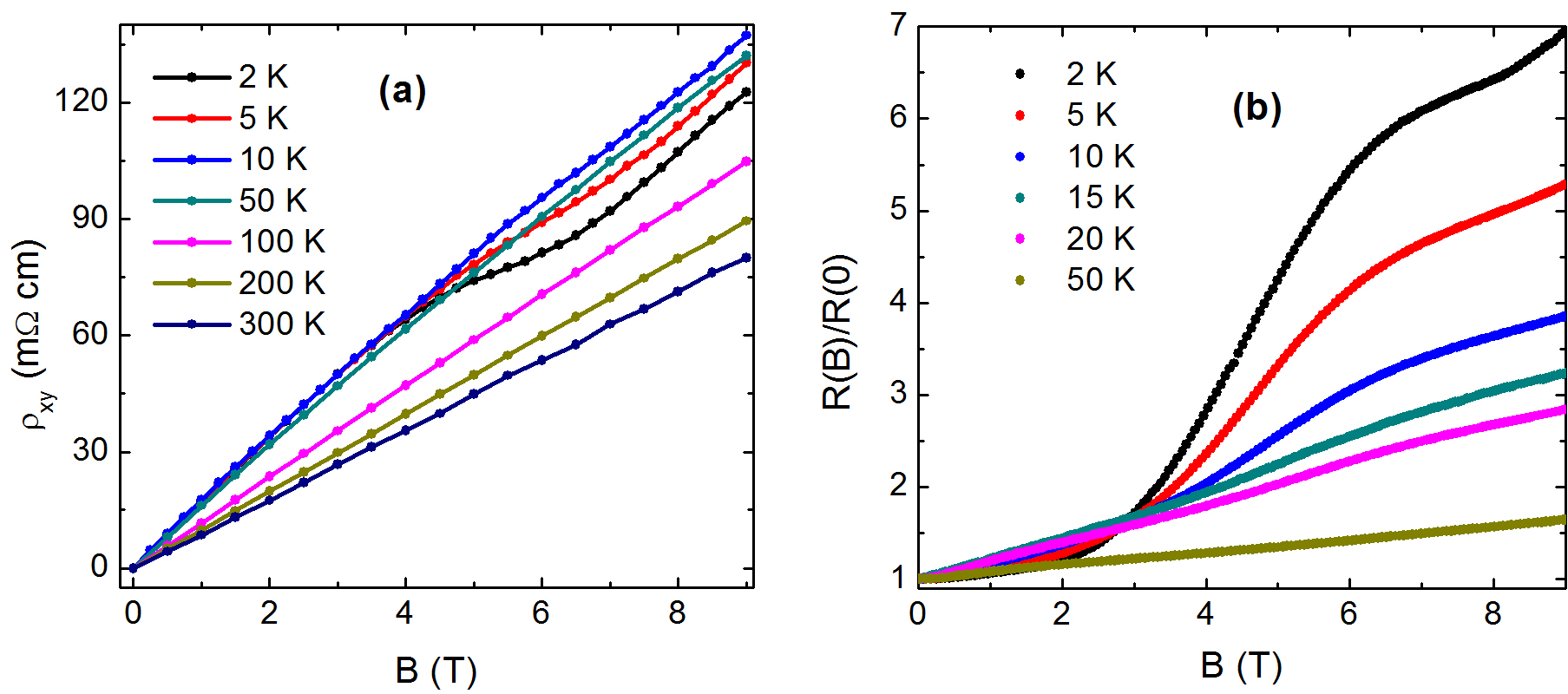}
\caption{(Color online) (a) Hall resistivity ($\rho$$_{xy}$) of ZrTe$_{5}$ single crystal at several representative temperatures over the range 2-300 K. (b) Magnetoresistance normalized to the zero field value upto 9 T.}\label{fig2}
\end{figure}
{\textbf{Longitudinal magnetoresistance and Chiral anomaly.}
As proposed by Hermann Weyl in 1929, the four-component massless Dirac equation in three spatial dimensions can be separated into two two-component equations, $\imath \frac{\partial \Psi}{\partial t}$=$\pm c\vec{\sigma}.\vec{p} \Psi$, where $\vec{\sigma}$ and $\vec{p}$ are the Pauli matrices and momentum respectively. The above equation describes particles with a definite chirality $\vec{\sigma}$.$\hat{p}$, known as Weyl fermions. Also, according to the classical equation of motion the number of fermions with plus or minus chirality is conserved separately.  The relativistic theory of charged chiral fermions in three spatial dimensions holds the so-called chiral anomaly$-$ non-conservation of chiral charge induced by external gauge fields with non-trivial topology, known as Adler-Bell-Jackiw anomaly  [27, 28]. Nielsen and Ninomiya  provided a physical picture of the chiral anomaly in the context of condensed matter physics [29]. Considering a band structure in three-dimensions which supports two Weyl nodes with opposite chirality separated in momentum space and applying a magnetic field along the line joining the Weyl nodes, they predicted an enhanced magneto-conductance due to the charge pumping between the two nodes in presence of an electric field ($\vec{E}$) parallel to $\vec{B}$. The application of magnetic field in three-dimensional Dirac semimetal splits the four-fold degenerate Dirac node into two Weyl nodes, along the direction of magnetic field [30, 31]. Initially, the plus or minus chirality fermions in the different Weyl nodes have same chemical potential $\mu$$^{+}$=$\mu$$^{-}$. Whereas, $\vec{E}$ parallel to $\vec{B}$ creates an imbalance ($\mu$$^{+}$$\neq$$\mu$$^{-}$) between the two Weyl nodes with opposite chirality, which induces a charge pumping from one Weyl node to another. The net current generation due to the chiral imbalance is j$_{c}$=$\frac{e^2B}{4\pi^2\hbar^2c}$($\mu$$^{+}$-$\mu$$^{-}$) [4, 31]. Again, ($\mu$$^{+}$-$\mu$$^{-}$) is proportional to $\vec{E}$.$\vec{B}$. As a result, the enhanced magneto-conductance is expected to show quadratic  $B$ dependence. Thus, the longitudinal magneto-conductance can be fitted with $\sigma$$_{c}$=$\sigma$$_{0}$+$a(T)B^{2}$, where $\sigma$$_{0}$ is the the zero-field conductivity. The field independent constant, $a(T)$ has the inverse $T$$^{2}$ dependence,
\begin{equation}
a(T)=\frac{e^2}{\pi\hbar}\frac{3}{8}\frac{e^2}{\hbar c}\frac{\nu^2}{\pi^3}\frac{\tau_{\upsilon}}{T^2+\frac{\mu^2}{\pi^2}},
\end{equation}
where $\nu$, $\tau_{\upsilon}$ and $\mu$ are the Fermi velocity, chirality changing scattering time and chemical potential respectively [4].\\
\begin{figure}
\includegraphics[width=1.0\textwidth]{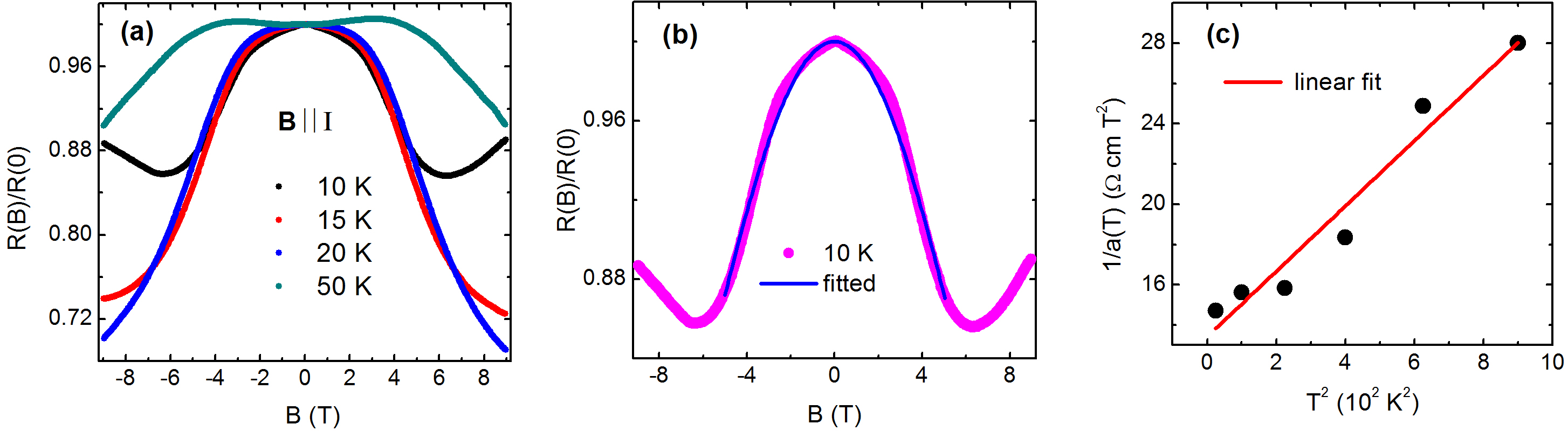}
\caption{(Color online) (a) Magnetoresistance measured at temperatures from 10 to 50 K, when applied current and magnetic field are parallel to each other, (b) Magnetoresistance at 10 K fitted with the theoretical expression, $\rho$$_{c}$=$\frac{1}{\sigma_{0}+a(T)B^{2}}$, (c) Temperature dependence of 1/$a$, where $a$ is in the units of S cm$^{-1}$ T$^{-2}$.}\label{fig3}
\end{figure}

To probe the  chiral anomaly, we have measured longitudinal magnetoresistance (LMR) by applying both the current and magnetic field along the \textbf{a} axis. As shown in Fig. 3 (a), the resistance at 10 K gradually decreases with increasing field until an upturn  occurs at high field. A small positive transverse MR component due to unavoidable misalignment in parallel configuration is responsible for this high field upturn. As the positive MR component rapidly suppresses with increasing temperature, the longitudinal negative MR becomes more clearly visible at higher temperatures. The LMR at several other temperatures from 5 to 40 K, has been shown in Supplementary Figure 2. With further increase in temperature above 40 K, however, the negative MR itself becomes very weak. The negative LMR was reproduced by several independent measurements and also verified in different crystals. LMR has been fitted with the above mentioned theoretical expression of the enhanced magneto-conductance and shown at a representative temperature 10 K in Figure 3(b). A good fitting between the theoretical expression and experimental data is reflected in above mentioned figure. By fitting LMR at different temperatures within 5 to 30 K, we have calculated the values of the parameter $a$. Observed value of $a(T)$ indicates that the strength of induced chiral current in the present case is comparable with the earlier report on ZrTe$_{5}$ [4]. In Figure 3(c), we have plotted $a$$^{-1}$ vs $T$$^{2}$. One can see from the figure that $a$$^{-1}$ is almost linear in $T^{2}$, as predicted theoretically. Thus, the negative longitudinal MR in $\vec{E}$$\parallel$$\vec{B}$ configuration implies four-component massless Dirac fermionic excitation in the bulk state of ZrTe$_{5}$ single crystal [4]. It may be noted that the negative MR due to induced chiral anomaly is a well established phenomenon in three-dimensional Dirac semimetals, which has also been observed in Cd$_{3}$As$_{2}$ [31] and Na$_{3}$Bi [32]. Presence of a gap between upper and lower Dirac cone in bulk may reduce the magnitude of the chiral current, but cannot destroy it fully [4]. \\
\begin{figure}
\includegraphics[width=0.8\textwidth]{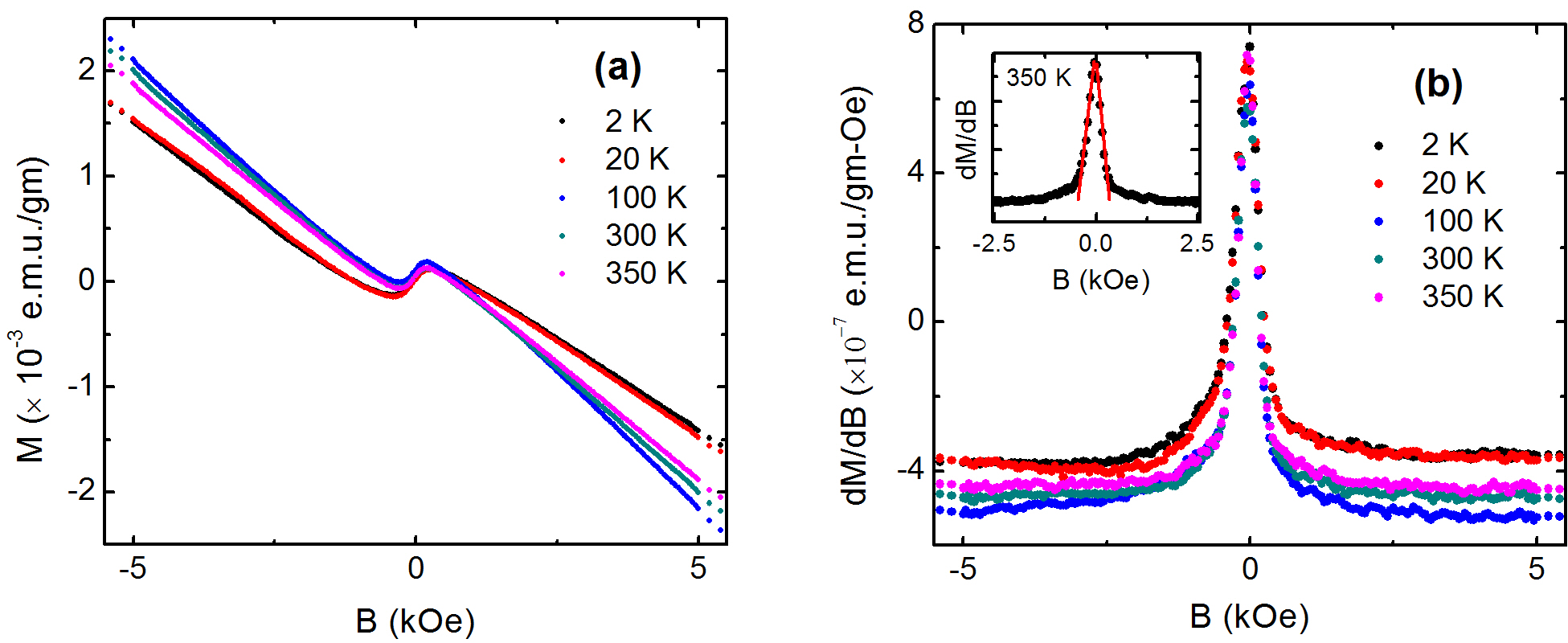}
\caption{(Color online) (a) Magnetization of ZrTe$_{5}$ single crystal, measured at several representative temperatures from 2 to 350 K. (b) Differential susceptibility ($\chi$=$\frac{dM}{dB}$) obtained after taking derivative of the magnetization with respect to external magnetic field. Inset shows the linear $B$ dependence of $\chi$ as $B$ tending towards zero on both side of the zero field cusp at representative temperature 350 K.}\label{fig3}
\end{figure}
{\textbf{Signature of topological surface state in magnetization measurements. }
The low-energy physics of the surface states for a three-dimensional topological insulator can be described by the Dirac type effective Hamiltonian, H$_{sur}$($k$$_{x}$,$k$$_{y}$)=$\hbar$$v$$_{F}$($\sigma$$^{x}$$k$$_{y}$-$\sigma$$^{y}$$k$$_{x}$), where $v$$_{F}$ is the Fermi velocity [33, 34].  Thus, for a fixed translational momentum $\vec{k}$, the ``spin", denoted by the Pauli matrix $\vec{\sigma}$, has a fixed direction for the eigenstate of the Hamiltonian. This is the most important property of the nontrivial topological surface states in 3D topological insulators, known as ``spin-momentum locking". As the ``spin" is always perpendicular to the momentum, one can introduce a helicity operator for the spin texture on circular constant energy contour of the Dirac cones [34], $\hat{h}$=(1/$k$)$\hat{z}$.($\vec{k}$$\times$$\vec{\sigma}$). This leads to left-handed spin texture for the upper Dirac cone and right-handed spin texture for the lower Dirac cone in the momentum space. Whereas at the Dirac point, as long as the Dirac spectrum is not gapped, the electron spin should be free to align along the applied magnetic field due to singularity in spin orientation [35]. This predicts  a low-field paramagnetic peak in the susceptibility  curve $\chi$($H$).\\

Figure 4(a) shows the magnetization of single crystal of diamagnetic ZrTe$_{5}$ with magnetic field along the \textbf{a} axis. Over the whole range of temperature from 2 to 350 K, ZrTe$_{5}$ shows diamagnetic signal except a paramagnetic upturn at low field region. It might be worthy to mention that single crystals of standard diamagnetic bismuth and  three-dimensional Dirac semimetal Cd$_{3}$As$_{2}$ do not show this type of behaviour [see Supplementary Figure 3 and Supplementary Figure 4(b)]. Figure 4(b) shows that a  cusp-like paramagnetic susceptibility  sharply rises above the diamagnetic floor in a narrow field range of $\sim$2 kOe around zero field. The height of the peak from the temperature dependent diamagnetic floor and its sharpness are insensitive to the temperature. This singular response of susceptibility shows no sign of thermal rounding up to as high as 350 K ($\sim$32 meV), which is almost one-half of the bulk band gap [4, 21].  Similar robust and singular paramagnetic response have been reported for the family of  three-dimensional topological insulators  which is the fingerprint of the helical spin texture of the topological Dirac fermions on the surface [35, 36]. Setting both the chemical potential $\mu$ and temperature to zero, one can show that this paramagnetic Dirac susceptibility decays linearly from its zero-field value at low field [35] as,
$\chi_{D} (B)\cong\frac{\mu_{0}}{4\pi^2}[\frac{(g\mu_{B})^2}{\hbar v_{F}}\Lambda-\frac{(g\mu_{B})^3}{\hbar^2 v_{F}^2}|B|+\ldots$]. Where $g$ is the Land\'{e} $g$-factor and $\Lambda$ is the effective size of the momentum space contributing to the singular part of free energy. It has been argued [35] that the maximum of the susceptibility, i.e., the peak height at zero field over the temperature dependent diamagnetic floor, depends on $\Lambda$, and thus may be controlled by the bulk bands. Whereas the nature of the singularity (i.e. cuspiness and linear-in-field decay of susceptibility at low fields, almost absence of thermal smearing, etc.,) is universal to the entire family of 3D topological insulators. Inset of Figure 4(b) shows the linear fit to the experimental data on the both sides of the zero-field cusp. The linear-in-field decay of the paramagnetic response, even at the highest measuring temperature 350 K, is evident from the figure. Often, surface states show ageing effect, which has been attributed to surface reconstruction and the formation of two-dimensional electron gas due to the bending of the bulk band at the surfaces [35, 37, 38]. Whether such behaviour is visible in the present case, similar measurements have been done on the same pieces of single crystals after three weeks of first measurement [See supplementary Figure 5]. Although the nature of the peak and its robustness against temperature remain unaffected, the reduction in peak height over time may be attributed to the expected ageing effect, similar to that observed in Bi$_{2}$Se$_{3}$, Sb$_{2}$Te$_{3}$ and  Bi$_{2}$Te$_{3}$ [35]. It has been pointed out that the spin/orbit texture may also exist in the bulk state of the material with strong spin-orbit coupling, such as in BiTeI [39] and WTe$_{2}$ [40]. Keeping this information in mind, one may think the possibility of the singular paramagnetic response from the bulk of ZrTe$_{5}$. But, as reported by the earlier ARPES measurements [14, 4, 21], the bulk state of ZrTe$_{5}$ holds a band gap ($\sim$50$\pm$10 meV) between the upper and lower Dirac cone, which disobeys the primary condition for the singularity in electron spin orientation from the spin/orbit texture. Secondly, the negative longitudinal magnetoresistance due to chiral charge imbalance under non-trivial gauge field and ARPES results, established the presence of four-component massless Dirac fermion in bulk 3D Dirac cone state of ZrTe$_{5}$. As far as we know, a four-component 3D Dirac fermion originating from spin-degenerate band, cannot have any spin-orbit texture. The age dependent reduction of the peak height, whereas the diamagnetic back ground is unaffected, also confirms the surface origin of this singular paramagnetic response.\\

{\textbf{\large Discussion}}\\
\begin{figure}[h!]
\includegraphics[width=0.5\textwidth]{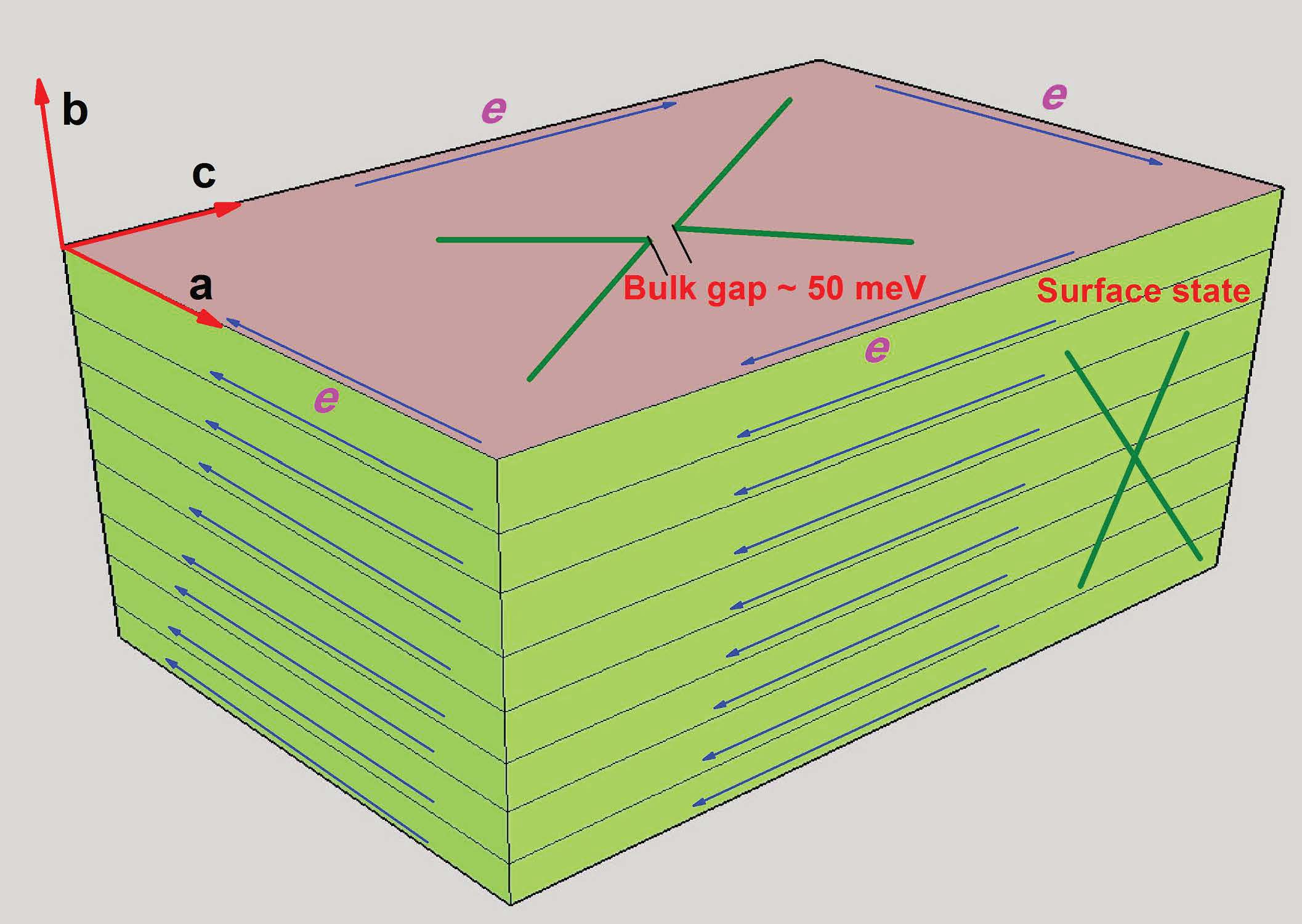}
\caption{(Color online) A schematic diagram, representing the minimum nontrivial topological nature of the electronic band structure of ZrTe$_{5}$ single crystal. ZrTe$_{5}$ monolayers, which lie in the \textbf{ac}-plane, stack together along the \textbf{b}-axis by weak van der Waals attraction. Independent conducting edge state of monolayers, as shown by the blue arrows in the figure, forms a topological surface state in bulk sample, known as weak three-dimensional topological insulating state.}\label{fig3}
\end{figure}
We have detected a robust zero-field paramagnetic peak in susceptibility due to the helical spin texture associated with the Dirac fermions of the surface state of three-dimensional topological insulators. Also, the negative longitudinal magnetoresistance implies induced chiral anomaly in ZrTe$_{5}$, which is the signature of three-dimensional Dirac fermion in the bulk. This allows one to conclude that ZrTe$_{5}$ is a novel quantum phase of matter, which hosts both topological  Dirac fermions on the surface and three-dimensional Dirac cone state with a mass gap between valence and conduction bands in the bulk. As mentioned earlier, ZrTe$_{5}$ can be either a weak or a strong three-dimensional topological insulator depending on the values of lattice parameters [12]. The simplest kind of topological insulators in three-dimension can be understood by stacking layers of the 2D quantum spin Hall insulator with weak van der Waals bonding between them [41], similar to that staking of monolayer in ZrTe$_{5}$ along \textbf{b}-axis. A schematic diagram has been shown in Figure 5, representing the possible minimum nontrivial electronic state of ZrTe$_{5}$ single crystal. Blue arrows represents conducting edge state of monolayer, which together with forms a topological surface state in bulk sample. On the other hand, the small gapped state in bulk is linear enough to show the signature of massless Diarc fermionic excitation in electronic transport. To the best of our knowledge, there is no such topological insulator with Dirac fermionic excitation in the bulk gap state in the history of material science. In addition, from the resistivity and Hall measurements in the present work, it is apparent that the anomalous peak in resistivity and its associated switching of charge carriers can be shifted to very low temperature.\\

\emph{Note added}. After completion of our work [arXiv:1603.05175] , we have come to know the results of Ref. [19]. From the scanning tunneling microscopy and angle-resolved photoemission spectroscopy, it has been shown that monolayer of ZrTe$_{5}$ is a large gap 2D topological insulator, as the  theory proposed [12]. The authors have found conducting state at monolayer step edge of \textbf{ac}-plane and far away from the edge i.e., well inside the plane, the spectrum is fully gapped. This is consistence with our proposed nontrivial electronic state for bulk ZrTe$_{5}$, as shown in schematic Figure 5. Very recently our results have been confirmed by Manzoni \emph{et al.} through ARPES [42].\\

{\textbf{\large Methods}}\\
 A stoichiometric mixture of the Zr (Alfa Aesar 99.9\%) and Te (Alfa Aesar 99.99\%) was sealed in a 15 cm long quartz tube with iodine ($\sim$ 5 mg/cc) and placed in a box furnace. It was then heated for seven days at 520$^{\circ}$C and cooled to room temperature at 10$^{\circ}$C/h. Next, the tube was shifted to a two-zone gradient furnace. One end of the tube containing the product was placed at 540$^{\circ}$C while the other end of the tube is placed at the cooler end of the furnace at 450 $^{\circ}$C to provide a temperature gradient for four days. After slowly cooling ($\sim$30$^{\circ}$C/h) it to room temperature, single crystals with needlelike morphology were obtained at the cooler end.\\
The resistivity measurements were done by standard four-probe technique. Electrical contacts were made using conductive silver paste. The electrical transport measurements were carried out in 9 T physical property measurement system (Quantum Design) and cryogen free measurement system (Cryogenic). The magnetization was measured using a Superconducting Quantum Interference Device$-$Vibrating Sample Magnetometer (SQUID-VSM) (Quantum Design). Before the magnetization measurements, the system was standardized using single crystal of diamagnetic bismuth (Alfa Aesar 99.99\%) and paramagnetic palladium [see Supplementary Figure 3 and Supplementary Figure 4(a)].\\

\textbf{Acknowledgement:} We thank V.A. Kulbachinskii for useful discussions.\\

\textbf{Author Contributions:}  A.P. prepared the sample. A.P. performed the experiments. A.P. and P.M. analysed and interpreted the data. A.P. and P.M. wrote the paper. P.M. supervised the project.\\

{\textbf{\large References}}\\

1. Furuseth, S., Brattas, L., \& Kjekshus, A. The Crystal Structure of HfTe$_{5}$. Acta Chem. Scand. \textbf{27}, 2367-2374 (1973).

2. Skelton, E.F., Wieting, T.J., Wolf, S.A., Fuller, W.W., Gubser, D.U., Francavilla, T.L. \& Levy, F. Giant resistivity and X-ray diffraction anomalies in low-dimensional ZrTe$_{5}$ and HfTe$_{5}$. Solid State Communications, \textbf{42}, 1-3 (1982).

3. Okada, S., Sambongi, T., \& Ido, M. Giant Resistivity Anomaly in ZrTe$_{5}$. J. Phys. Soc. Jpn. \textbf{49}, 839-840 (1980).

4. Li, Q., Kharzeev, D. E., Zhang, C., Huang, Y., Pletikosić, I., Fedorov, A. V., Zhong,	R. D., Schneeloch, J. A., Gu, G. D., \& Valla, T. Chiral magnetic effect in ZrTe$_{5}$. Nat. Phys. \textbf{12}, 550-554 (2016).

5. Chen, R. Y., Zhang, S. J., Schneeloch, J. A., Zhang, C., Li, Q., Gu, G. D. \& Wang, N. L. Optical spectroscopy study of the three-dimensional Dirac semimetal ZrTe$_{5}$. Phys. Rev. B \textbf{92}, 075107 (2015).

6. Jones, T. E., Fuller, W. W., Wieting, T. J. \& Levy, F. Thermoelectric power of  HfTe$_{5}$ and ZrTe$_{5}$. Solid State Commun. \textbf{42}, 793-798 (1982).

7. Izumi, M., Uchinokura, K., Matsuura, E. \& Harada, S. Hall effect and transverse magnetoresistance in a low-dimensional conductor HfTe$_{5}$. Solid State Commun. \textbf{42}, 773-778 (1982).

8. Ong, N. P. \& Monceau, P. Anomalous transport properties of a linear-chain metal: NbSe$_{3}$. Phys. Rev. B \textbf{16}, 3443 (1977).

9. Okada, S., Sambongi, T., Ido, M., Tazuke, Y., Aoki, R. \& Fujita, O. Negative Evidences for Charge/Spin Density Wave in ZrTe$_{5}$. J. Phys. Soc. Jpn. \textbf{51}, 460-467 (1982).

10. Rubinstein, M. HfTe$_{5}$ and ZrTe$_{5}$ : Possible polaronic conductors. Phys. Rev. B \textbf{60}, 1627 (1999).

11. McIlroy, D. N., Moore, S., Zhang, D., Wharton, J., Kempton, B., Littleton, R., Wilson, M., Tritt, T. M., \& Olson, C. G. Observation of a semimetal–semiconductor phase transition in the intermetallic ZrTe$_{5}$. J. Phys.: Condens. Matter \textbf{16}, L359–L365 (2004).

12. Weng, H., Dai, X., \& Fang, Z. Transition-Metal Pentatelluride ZrTe$_{5}$ and HfTe$_{5}$: A Paradigm for Large-Gap Quantum Spin Hall Insulators. Phys. Rev. X \textbf{4}, 011002 (2014).

13. Tritt, T. M., Lowhorn, N. D., Littleton IV, R. T., Pope, A., Feger, C. R. \& Kolis, J. W. Large enhancement of the resistive anomaly in the pentatelluride materials HfTe$_{5}$ and ZrTe$_{5}$ with applied magnetic field. Phys. Rev. B \textbf{60}, 7816 (1999).

14. Manzoni, G., Sterzi, A., Crepaldi, A., Diego, M., Cilento, F., Zacchigna, M., Bugnon, Ph., Berger, H., Magrez, A., Grioni, M. \& Parmigiani, F. Ultrafast Optical Control of the Electronic Properties of ZrTe$_{5}$. Phys. Rev. Lett. \textbf{115}, 207402 (2015).

15. Chen, R. Y., Chen, Z. G., Song, X.-Y., Schneeloch, J. A., Gu, G. D., Wang, F. \& Wang, N. L. Magnetoinfrared Spectroscopy of Landau Levels and Zeeman Splitting of Three-Dimensional Massless Dirac Fermions in ZrTe$_{5}$. Phys. Rev. Lett. \textbf{115}, 176404 (2015).

16. Zhou, Y., Wu, J., Ning, W., Li, N., Du, Y., Chen, X., Zhang, R., Chi, Z., Wang, X., Zhu, X., Lu, P., Ji, C., Wan, X., Yang, Z., Sun, J., Yang, W., Tian, M. \& Zhang, Y. Pressure-induced semimetal to superconductor transition in a three-dimensional topological material ZrTe$_{5}$. Proc. Natl. Acad. Sci. USA \textbf{113}(11), 2904-2909 (2016).

17. Yuan, X., Zhang, C., Liu, Y., Song, C., Shen, S., Sui, X., Xu, J., Yu, H., An, Z., Zhao, J., Yan, H. \& Xiu, F. Observation of quasi-two-dimensional Dirac fermions in ZrTe$_{5}$. arXiv:1510.00907.

18. Niu, J., Wang, J., He, Z., Zhang, C., Li, X., Cai, T., Ma, X., Jia, S., Yu, D. \& Wu, X. Electrical transport in nano-thick ZrTe$_{5}$ sheets: from three to two dimensions. arXiv:1511.09315.

19. Wu, R., Ma, J.-Z., Zhao, L.-X., Nie, S.-M., Huang, X., Yin, J.-X., Fu, B.-B., Richard, P., Chen, G.-F., Fang, Z., Dai, X., Weng, H.-M., Qian, T., Ding, H. \& Pan, S. H. Experimental evidence of large-gap two-dimensional topological insulator on the surface of ZrTe$_{5}$. Phys. Rev. X \textbf{6}, 021017 (2016).

20. Yu, W., Jiang, Y., Yang, J., Dun, Z. L., Zhou, H. D., Jiang, Z., Lu, P. \& Pan, W. Quantum Oscillations at Integer and Fractional Landau Level Indices in ZrTe$_{5}$. 	arXiv:1602.06824.

21. Zhang, Y. \emph{et al.} Electronic Evidence of Temperature-Induced Lifshitz Transition and Topological nature in ZrTe$_{5}$. arXiv:1602.03576.

22. DiSalvo, F. J., Fleming, R. M., \& Waszczak, J. V. Possible phase transition in the quasi-one-dimensional materials ZrTe$_{5}$ or HfTe$_{5}$. Phys. Rev. B \textbf{24}, 2935 (1981).

23. Fjellvåg, H. \& Kjekshus, A. Structural properties of ZrTe$_{5}$ and HfTe$_{5}$ as seen by powder diffraction. Solid State Commun. \textbf{60}, 91-93 (1986).

24. Okada, S., Sambongi, T. \& Ido, M. Giant Resistivity Anomaly in ZrTe$_{5}$. J. Phys. Soc. Jpn. \textbf{49}, 839 (1980).

25. Littleton IV, R. T., Tritt, T. M., Kolis, J. W., Ketchum, D. R., Lowhorn, N. D. \& Korzenski, M. B. Suppression of the resistivity anomaly and corresponding thermopower behavior in the pentatelluride system by the addition of Sb: Hf$_{1-x}$Zr$_{x}$Te$_{5-y}$Sb$_{y}$. Phys. Rev. B \textbf{64}, 121104(R) (2001).

26. Izumi, M., Nakayama, T., Uchinokura, K., Harada, S., Yoshizaki, R. \& Matsuura, E. Shubnikov-de Haas oscillations and Fermi surfaces in transition-metal pentatellurides ZrTe$_{5}$ and HfTe$_{5}$. J. Phys. C: Solid State Phys. \textbf{20}, 3691 (1987).

27. Adler, S. L. Axial-Vector Vertex in Spinor Electrodynamics. Phys. Rev. \textbf{177}, 2426 (1969).

28. Bell, J. S. \& Jackiw, R., A PCAC puzzle: $\pi$$^{0}$$\longrightarrow$$\gamma$$\gamma$ in the $\sigma$-model. Il Nuovo Cimento A \textbf{60}, 47-61 (1969).

29. Nielsen, H. B. \& Ninomiya, M. The Adler-Bell-Jackiw anomaly and Weyl fermions in a crystal. Phys. Lett. B \textbf{130}, 389-396 (1983).

30. Gorbar, E. V., Miransky, V. A., \& Shovkovy, I. A. Engineering Weyl nodes in Dirac semimetals by a magnetic field. Phys. Rev. B \textbf{88}, 165105 (2013).

31. Li, C. Z., Wang, L. X., Liu, H., Wang, J., Liao, Z. M. \& Yu, D. P. Giant negative magnetoresistance induced by the chiral anomaly in individual Cd$_{3}$As$_{2}$ nanowires. Nat. Commun. \textbf{6}, 10137 (2015).

32. Xiong, J., Kushwaha, S. K., Liang, T., Krizan, J. W., Hirschberger, M., Wang, W., Cava, R. J. \& Ong, N. P. Evidence for the chiral anomaly in the Dirac semimetal Na$_{3}$Bi. Science \textbf{350}, 413 (2015).

33. Hsieh, D. \emph{et al.} A tunable topological insulator in the spin helical Dirac transport regime. Nature \textbf{460}, 1101 (2009).

34. Zhang, H., Liu, C. -X. \& Zhang, S. -C. Spin-Orbital Texture in Topological Insulators. Phys. Rev. Lett. \textbf{111}, 066801 (2013).

35. Zhao, L., Deng, H., Korzhovska, I., Chen, Z., Konczykowski, M., Hruban, A., Oganesyan, V. \& Krusin-Elbaum, L. Singular robust room-temperature spin response from topological Dirac fermions. Nat. Mater. \textbf{13}, 580 (2014).

36. Buga, S. G., Kulbachinskii, V. A., Kytin, V. G., Kytin, G. A., Kruglov, I. A., Lvova, N. A., Perov, N. S., Serebryanaya, N. R., Tarelkin, S. A. \& Blank, V. D. Superconductivity in bulk polycrystalline metastable phases of Sb$_{2}$Te$_{3}$ and Bi$_{2}$Te$_{3}$ quenched after high-pressure-high-temperature treatment. Chem. Phys. Lett. \textbf{631-632}, 97 (2015).

37. He, X., Zhou, W., Wang, Z. Y., Zhang, Y. N., Shi, J., Wu, R. Q. \& Yarmoff, J. A. Surface Termination of Cleaved Bi$_{2}$Se$_{3}$ Investigated by Low Energy Ion Scattering. Phys.Rev.Lett. \textbf{110}, 156101(2013).

38. Bahramy, M. S., King, P. D. C., delaTorre, A., Chang, J., Shi, M., Patthey, L., Balakrishnan, G., Hofmann, Ph., Arita, R., Nagaosa, N. \& Baumberger, F. Emergent quantum confinement at topological insulator surfaces. Nat. Commun. \textbf{3},1159(2012).

39. Murakawa, H., Bahramy, M. S., Tokunaga, M., Kohama, Y., Bell, C., Kaneko, Y., Nagaosa, N., Hwang, H. Y. \& Tokura, Y. Detection of Berry's Phase in a Bulk Rashba Semiconductor. Science \textbf{342}, 1490 (2013).

40. Jiang, J., Tang, F., Pan, X. C., Liu, H. M., Niu, X. H., Wang, Y. X., Xu, D. F., Yang, H. F., Xie, B. P., Song, F. Q., Dudin, P., Kim, T. K., Hoesch, M., Kumar Das, P., Vobornik, I., Wan, X. G. \& Feng, D. L. Signature of Strong Spin-Orbital Coupling in the Large Nonsaturating Magnetoresistance Material WTe$_{2}$. Phys. Rev. Lett. \textbf{115}, 166601 (2015).

41. Hasan, M. Z. \& Kane, C. L. Colloquium: Topological insulators. Rev. Mod. Phys. \textbf{82}, 3045 (2010).

42. Manzoni, G. \emph{et al.} Evidence for a Strong Topological Insulator Phase in ZrTe$_{5}$. arXiv:1608.03433.\\
\newpage
{\large Supplementary information for ``Coexistence of topological Dirac fermions in the surface and three-dimensional Dirac cone state in the bulk of ZrTe$_{5}$ single crystal"}

\subsection{\large Sample Characterization}
Phase purity and the structural analysis of the samples were done using the high-resolution powder x-ray diffraction (XRD) technique (Rigaku, TTRAX II) using Cu-K$_{\alpha}$ radiation. FIG.S1 shows the x-ray diffraction pattern of powdered sample of ZrTe$_{5}$ single crystals at room temperature. Within the resolution of XRD, we have not seen any peak due to the impurity phase.  Using the Rietveld profile refinement program of diffraction patterns, we have calculated the lattice parameters $a$=3.96 {\AA}, $b$=14.50 {\AA} and $c$$=$13.78 {\AA} with space group symmetry $Cmcm$.\\
\begin{figure}[h!]
% Requires \usepackage{graphicx}
\includegraphics[width=0.55\textwidth]{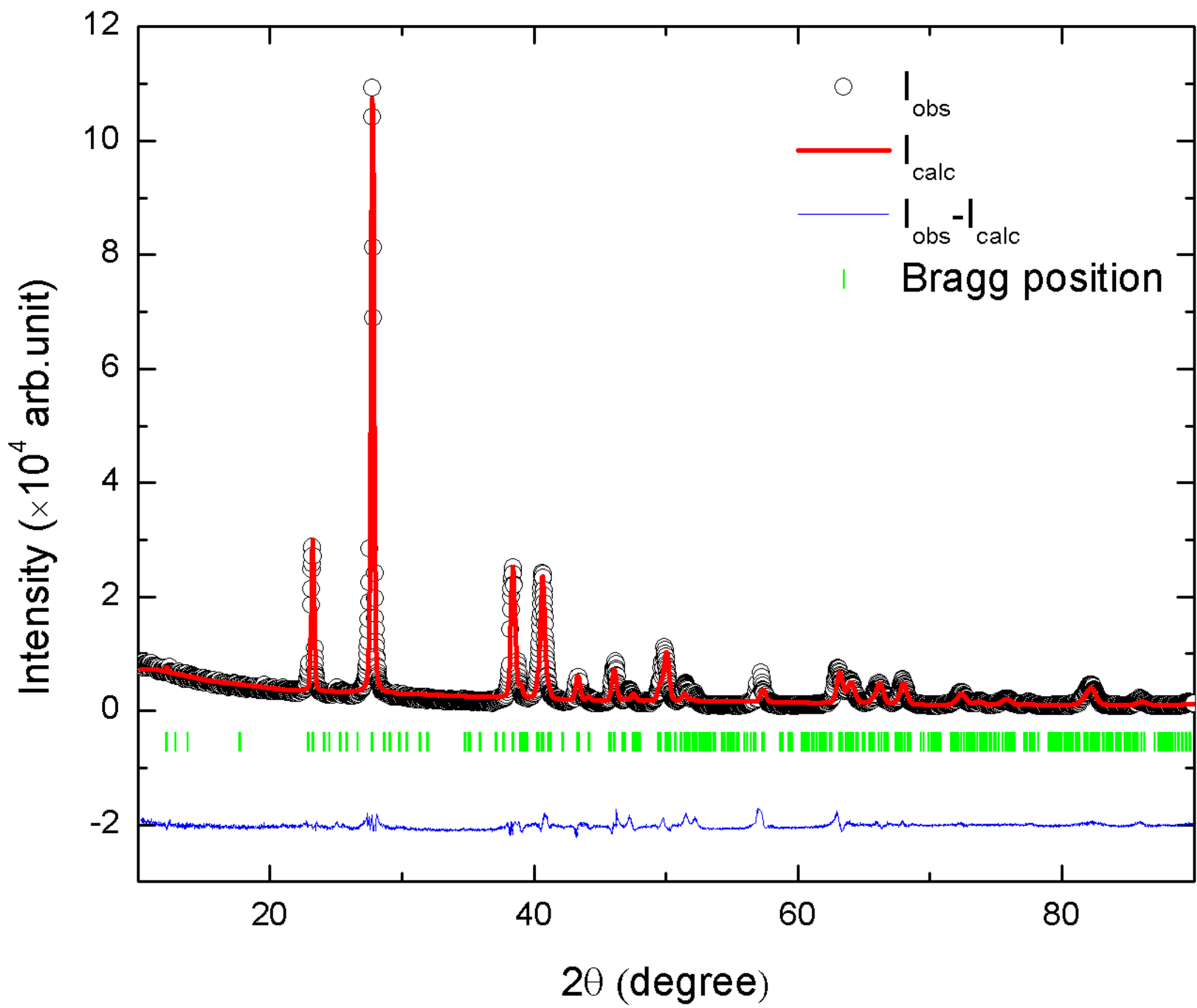}
\renewcommand{\figurename}{FIG.S}
\caption{(Color online) X-ray diffraction pattern of powdered single crystals of ZrTe$_{5}$. Black open circle, experimental data; red, the calculated pattern; blue, the difference between experimental and calculated intensities; green, the Bragg positions.}\label{rh}
\end{figure}
\subsection{\large Longitudinal magnetoresistance at several representative temperatures}
\begin{figure}[h!]
% Requires \usepackage{graphicx}
\includegraphics[width=0.55\textwidth]{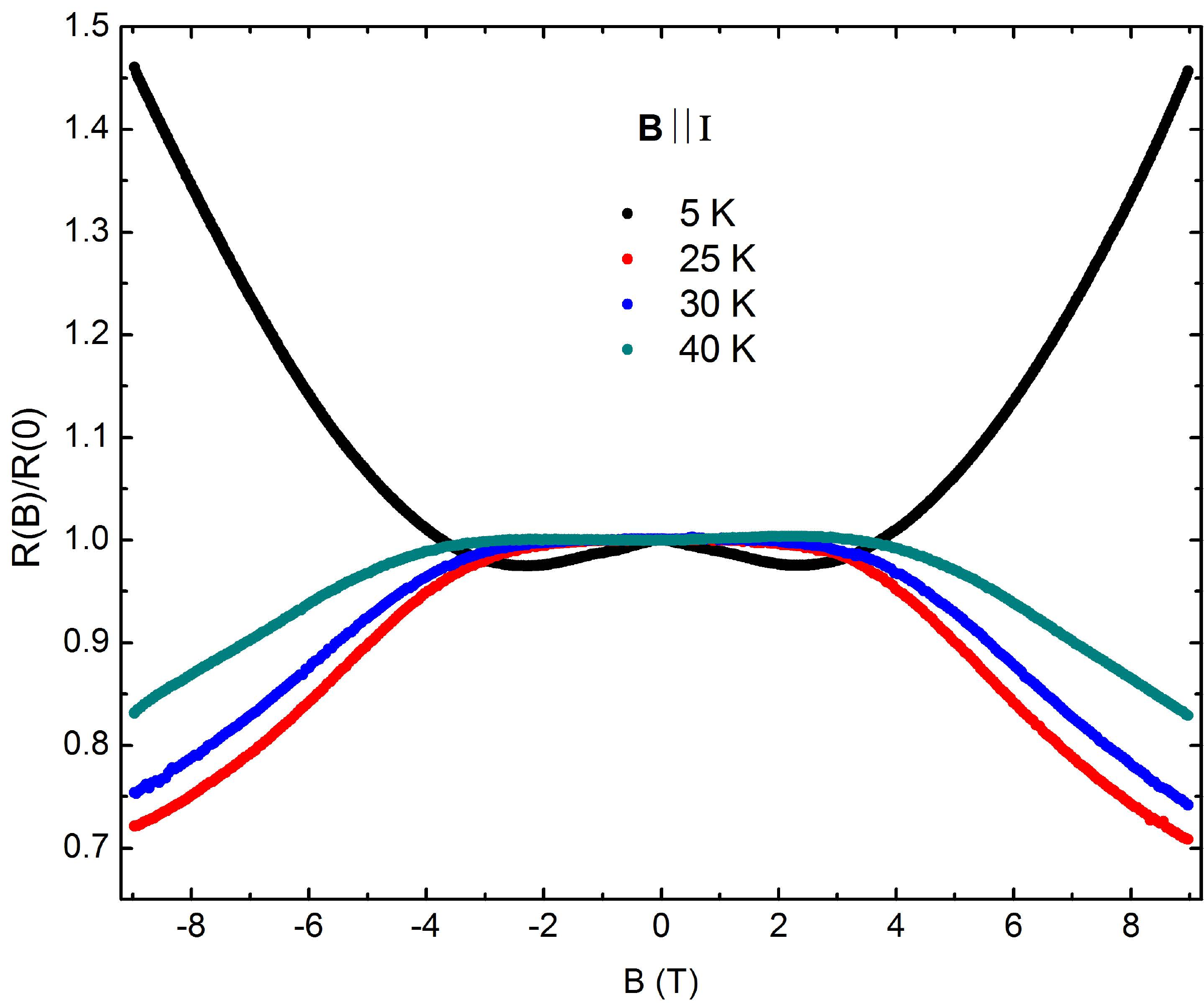}
\renewcommand{\figurename}{FIG.S}
\caption{(Color online) Magnetoresistance measured at temperatures 5 K, 25 K, 30 K and 40 K, when applied current and magnetic field are parallel to each other.}\label{rh}
\end{figure}
\pagebreak
\subsection{\large Magnetic measurements of standard samples}
We have done magnetization measurements of standard bismuth and palladium samples in SQUID-VSM [MPMS 3, Quantum Design] prior to ZrTe$_{5}$ single crystal. FIG.S3 (a) shows that linear diamagnetic moment of bismuth at 2, 100 and 300 K passes through the origin. This is more clear from FIG.S2 (b), which shows magnetic field dependence of differential susceptibility ($\chi$=$\frac{dM}{dB}$). With quantum oscillation at high fields, absence of any paramagnetic peak around zero field implies that singular paramagnetic susceptibility in ZrTe$_{5}$ single crystal is not due to any spurious response in our system.\\
\begin{figure}[h!]
% Requires \usepackage{graphicx}
\includegraphics[width=0.7\textwidth]{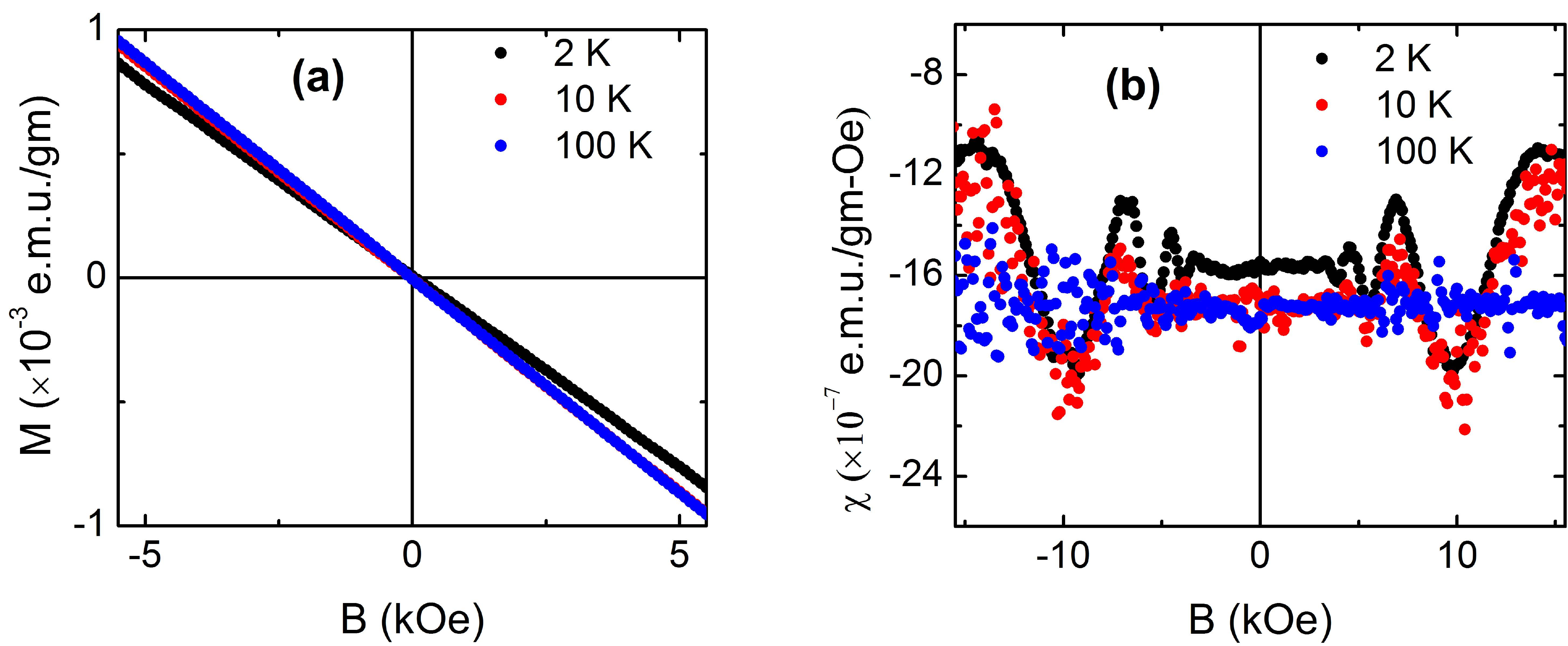}
\renewcommand{\figurename}{FIG.S}
\caption{(Color online) (a) Magnetization measured in SQUID-VSM (Quantum Design) at several representative temperatures for standard diamagnetic bismuth sample. (b) Differential susceptibility ($\chi$=$\frac{dM}{dB}$) obtained after taking numerical derivative of the magnetization with respect to external magnetic field.}\label{rh}
\end{figure}
FIG. S4(a) shows the expected magnetic behaviour of paramagnetic palladium sample, provided by the Quantum Design. FIG.S4(b) shows the low field susceptibility of palladium at 2 K and room temperature. The nonlinear behaviour of $\chi$ at low field and a broad zero field peak at 2 K are completely suppressed at room temperature. This is entirely different from the singular, robust and linear low field paramagnetic response from the topological surface state in ZrTe$_{5}$. For the sake of fundamental interest, we have also measured the magnetization of a single crystal of three-dimensional Dirac semimetal Cd$_{3}$As$_{2}$, as shown in FIG.S4(c). This shows perfect diamagnetic behaviour with no paramagnetic sign at low field.\\

\begin{figure}[h!]
% Requires \usepackage{graphicx}
\includegraphics[width=1.0\textwidth]{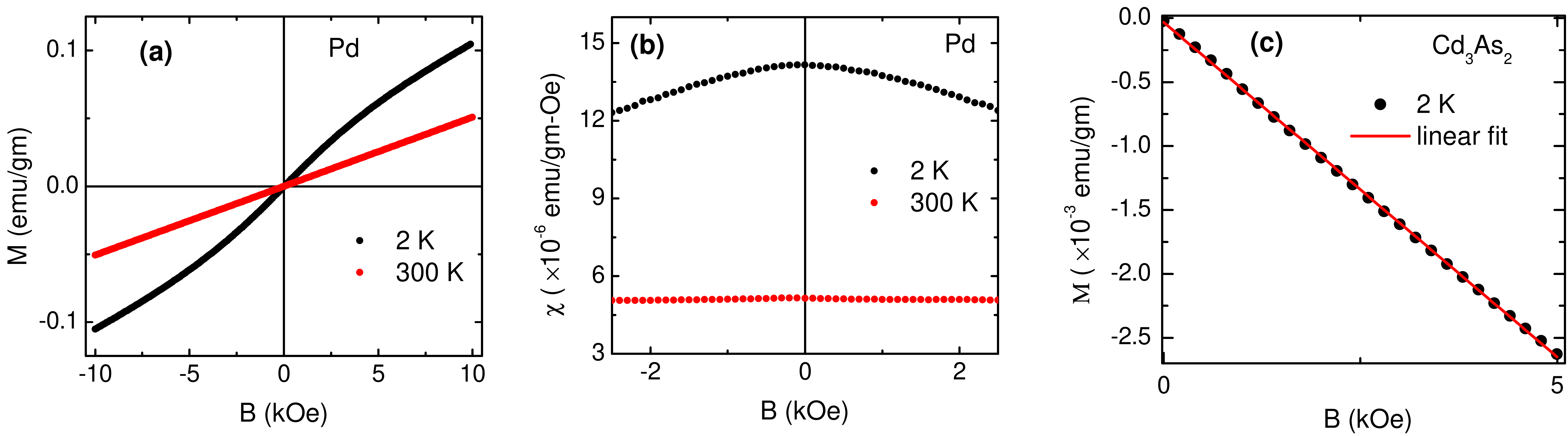}
\renewcommand{\figurename}{FIG.S}
\caption{(Color online) (a) Magnetization of standard paramagnetic palladium sample at 2 and 300 K, (b) Differential susceptibility ($\chi$=$\frac{dM}{dB}$) obtained after taking numerical derivative of the magnetization with respect to external magnetic field, and (c) Magnetization of single crystal of three-dimensional Dirac semimetal Cd$_{3}$As$_{2}$. Solid line implies linear fit to the experimental data.}\label{rh}
\end{figure}

\begin{figure}
\includegraphics[width=0.7\textwidth]{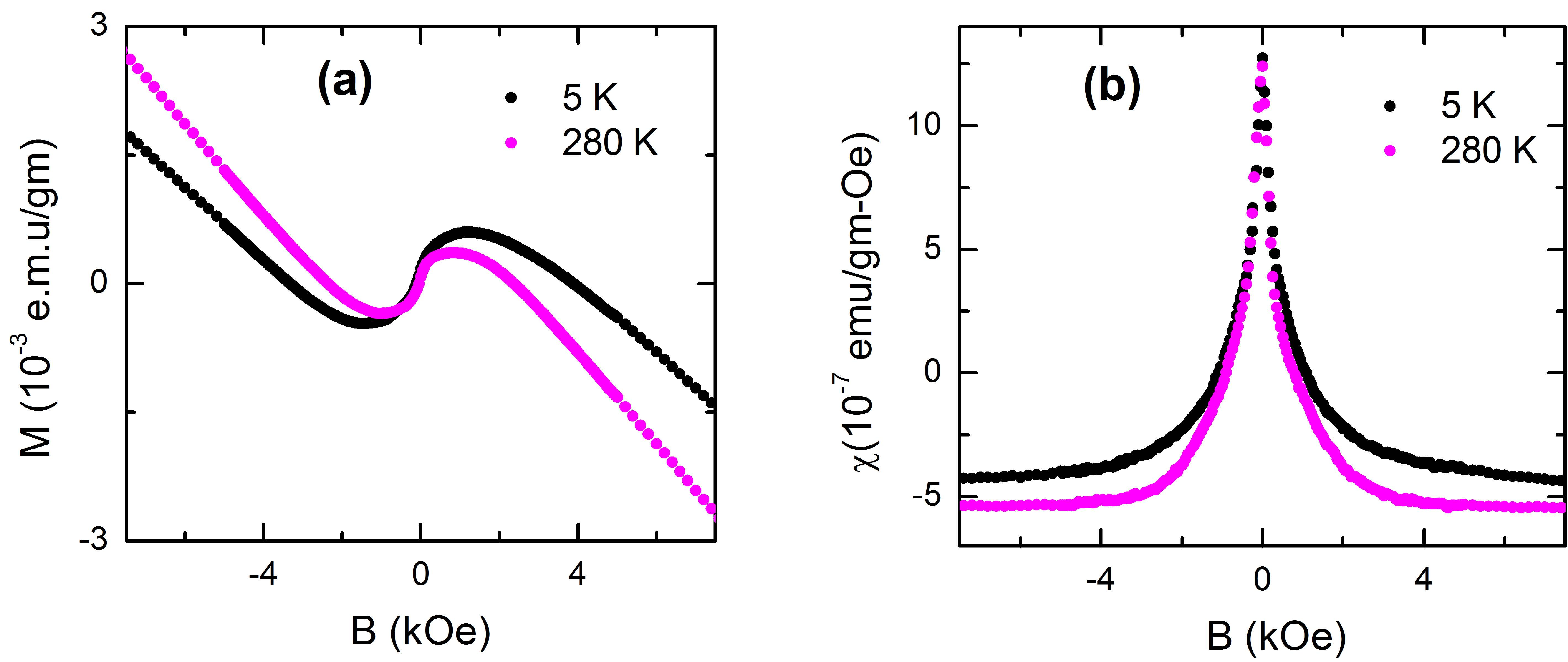}
\renewcommand{\figurename}{FIG.S}
\caption{(a) Magnetization (M) vs $B$ of freshly cleaved single crystal of Bi$_{1.5}$Sb$_{0.5}$Te$_{1.7}$Se$_{1.3}$ at temperatures 5 K and 280 K, (b) Susceptibility ($\chi$=$\frac{dM}{dB}$) as a function of $B$, calculated by taking the first order derivative of magnetization.}\label{rh}
\end{figure}
In this context, we would also like to mention that similar magnetization measurements have been performed on single crystal of well established three-dimensional topological insulator Bi$_{1.5}$Sb$_{0.5}$Te$_{1.7}$Se$_{1.3}$ [1,2] using the same experimental setup. We have observed singular robust paramagnetic susceptibility peak like to that observed in ZrTe$_{5}$ and shown in FIG.S5(a) and FIG.S5(b) at representative temperatures 5 K and 280 K. The details will be reported in another independent work.\\
\subsection{\large Aging effect of ZrTe$_{5}$ single crystal}
Figure S6 (a) shows the magnetization of ZrTe$_{5}$ single crystal after three weeks from the first measurements [Fig. 4(b)]. As shown in Fig. S6 (b), the singular paramagnetic response remains as robust as the earlier measurements [Fig. 4(b) and Fig. S6 (c)] against temperature. It also retains the similar linear in field decay on the both sides of the zero-field cusp, as shown in the inset of figure S6 (b) at a representative temperature 350 K. But the overall cusp height has been reduced over time. Similar reduction in peak height has been observed in Bi$_{2}$Se$_{3}$, Sb$_{2}$Te$_{3}$ and  Bi$_{2}$Te$_{3}$ [3] and has been attributed to surface reconstruction and the formation of two-dimensional electron gas due to bending of bulk band at the surfaces [3-5].
\pagebreak
\newpage
\begin{figure}[h!]
\includegraphics[width=1.0\textwidth]{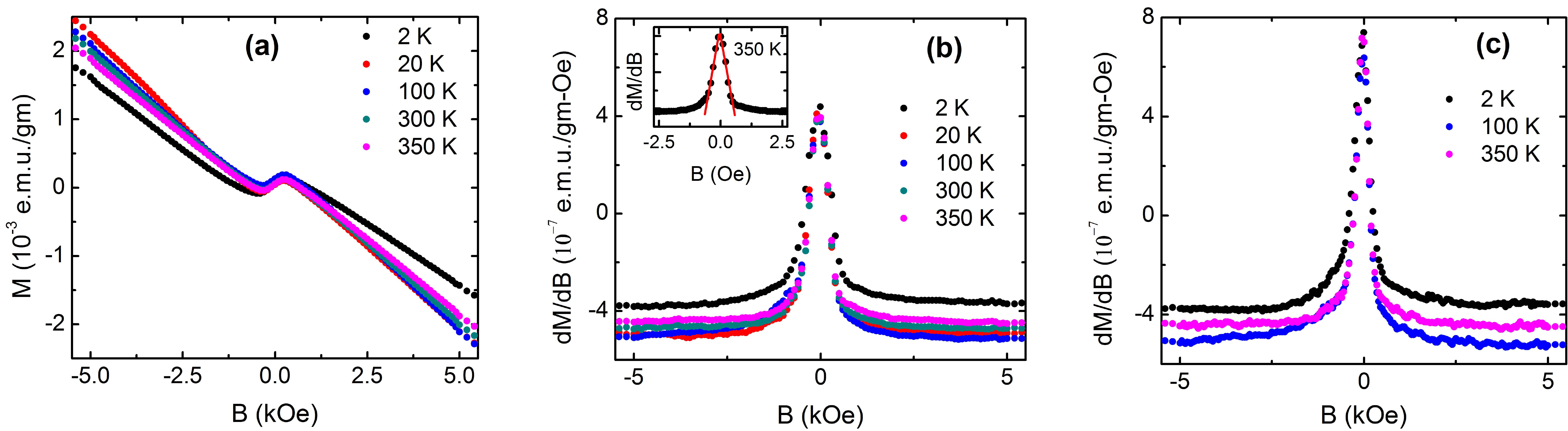}
\renewcommand{\figurename}{FIG.S}
\caption{(Color online) (a) Magnetization of ZrTe$_{5}$ single crystal, which was kept for three weeks in desiccator from the first measurements [Fig.4(b)]. (b) Differential susceptibility ($\chi$=$\frac{dM}{dB}$) obtained after taking derivative of the magnetization with respect to external magnetic field. Inset shows the linear $B$ dependence of $\chi$ as $B$ tending towards zero on both sides of the zero field cusp at representative temperature 350 K. (c) Susceptibility of the earlier measurements [Fig.4(b)] is shown to compare the nature of the peak.}\label{rh}
\end{figure}

{\textbf{\large References}}\\

1. Arakane, T., Sato, T., Souma, S., Kosaka, K., Nakayama, K., Komatsu, M., Takahashi, T., Ren, Z., Segawa, K. \& Ando, Y. Tunable Dirac cone in the topological insulator Bi$_{2-x}$Sb$_{x}$Te$_{3-y}$Se$_{y}$. Nat. Commun. \textbf{3}, 636 (2012).\\

2. Kim, S., Yoshizawa, S., Ishida, Y., Eto, K., Segawa, K., Ando, Y., Shin, S. \& Komori, F. Robust Protection from Backscattering in the Topological Insulator Bi$_{1.5}$Sb$_{0.5}$Te$_{1.7}$Se$_{1.3}$. Phys. Rev. Lett. \textbf{112}, 136802 (2014).\\

3. Zhao, L., Deng, H., Korzhovska, I., Chen, Z., Konczykowski, M., Hruban, A., Oganesyan, V. \& Krusin-Elbaum, L. Singular robust room-temperature spin response from topological Dirac fermions. Nat. Mater. \textbf{13}, 580 (2014).\\

4.He, X., Zhou, W., Wang, Z. Y., Zhang, Y. N., Shi, J., Wu, R. Q. \& Yarmoff, J. A. Surface Termination of Cleaved Bi$_{2}$Se$_{3}$ Investigated by Low Energy Ion Scattering. Phys.Rev.Lett. \textbf{110}, 156101(2013).\\

5. Bahramy, M. S., King, P. D. C., delaTorre, A., Chang, J., Shi, M., Patthey, L., Balakrishnan, G., Hofmann, Ph., Arita, R., Nagaosa, N. \& Baumberger, F. Emergent quantum confinement at topological insulator surfaces. Nat. Commun. \textbf{3},1159(2012).\\
\end{document}